\documentclass[slac_one]{revtex4}
\usepackage{graphicx}
\usepackage{fancyhdr}
\begin{document}

\title{Search for new physics in $B_s \rightarrow \mu^+ \mu^-$ and $B \rightarrow K^{(*)} \mu^+
  \mu^-$}

%

\author{N. Serra on behalf of the LHCb Collaboration}
\affiliation{Universit\"{a}t Z\"{u}rich, Z\"{u}rich, CH-8057, Switzerland}
%

\begin{abstract}
Transitions of the type $b \to s l^+ l^-$ are flavour changing neutral current
processes where new physics can enter in competing loop diagrams 
with respect to the  Standard Model contributions.  
In these decays several observables sensitive to new physics, and where theoretical uncertainties are under control, can be constructed. 
Particularly interesting are the angular asymmetries in the decay $B_d
\to K^* \mu^+ \mu^-$ and the measurement of the branching fraction of the
decays $B_{s,d} \to \mu^+ \mu^-$. 
Recent measurements of these observables and the measurement of the
isospin asymmetry in the decays  $B \to K^{(*)} \mu^+ \mu^-$ are
presented. 
\end{abstract}

\maketitle

\thispagestyle{fancy}


\section{INTRODUCTION} 
Rare decays which proceed via Flavour Changing Neutral Currents
(FCNC) are induced by one-loop diagrams in the Standard Model (SM) and are excellent
probes for New Physics (NP).  
New particles can enter in competing loop-order diagrams, resulting in
large deviations from SM predictions. 
In general, an effective Hamiltonian formalism is used to describe the amplitudes of
FCNC processes, according to the formula:
\begin{equation}
H_{eff} = \frac{G_F}{\sqrt{2}}\sum_i V_{CKM}^i C_i(\mu) Q_i\;,
\end{equation}   
where $V^i_{CKM}$ are the relevant factors of the CKM matrix; $Q_i$ are local operators; $C_i$ are the corresponding couplings
(Wilson coefficients);  and $\mu$ is the QCD renormalization scale.
The information on heavy degrees of freedom is embedded in the
Wilson coefficients, which are, in general, different for the SM and NP
models. 
Moreover, the correlation of different channels, where different combinations of
Wilson coefficients contribute, is a powerful tool for searching and
understanding the structure of NP. This approach is complementary to
the direct search for NP at general purpose detectors. At current
experiments, precision measurements of flavour physics allow energy scales
as high as 200 TeV~\cite{Buras:2009if} to be tested, which exceed the reach for direct production by about two
orders of magnitude\footnote{The exact value of the
  NP scale that can be tested with flavour physics observables is
  model dependent. }. In this paper recent measurements of the decays $B_{s,d}
\rightarrow \mu^+ \mu^-$ and $B \rightarrow K^{(*)} \mu^+ \mu^-$ and
the implication for the search for physics beyond the SM will be discussed.

\section{The $B_s \rightarrow \mu^+ \mu^-$ decay}

The decays $B_{d,s} \to \mu^+ \mu^-$ are suppressed in the SM, being FCNC and helicity suppressed. Their branching fractions are
predicted to be: $ {\cal B} (B_s\rightarrow \mu^{+} \mu^{-}) = (3.2 \pm 0.2)\times 10^{-9}$ and 
${\cal B}(B_d\rightarrow \mu^{+} \mu^{-})= (1.0 \pm 0.1)\times
10^{-10}$ in the SM~\cite{Buras:2010mh,Buras:2010wr}. 
It is worth emphasising that while the
theory predictions refer to branching fractions at $t=0$, experiments
measure time integrated branching fractions. This implies that the
effective measured branching fraction for $B_s \to \mu^+ \mu^-$ is about
10\% larger than predictions~\cite{deBruyn:2012wk,deBruyn:2012wj}. 
The correction due to soft photons, which are not detected
experimentally, is discussed in Ref.~\cite{Buras:2012ru}.

\begin{figure}[!h]
\centering
\includegraphics[scale=0.40]{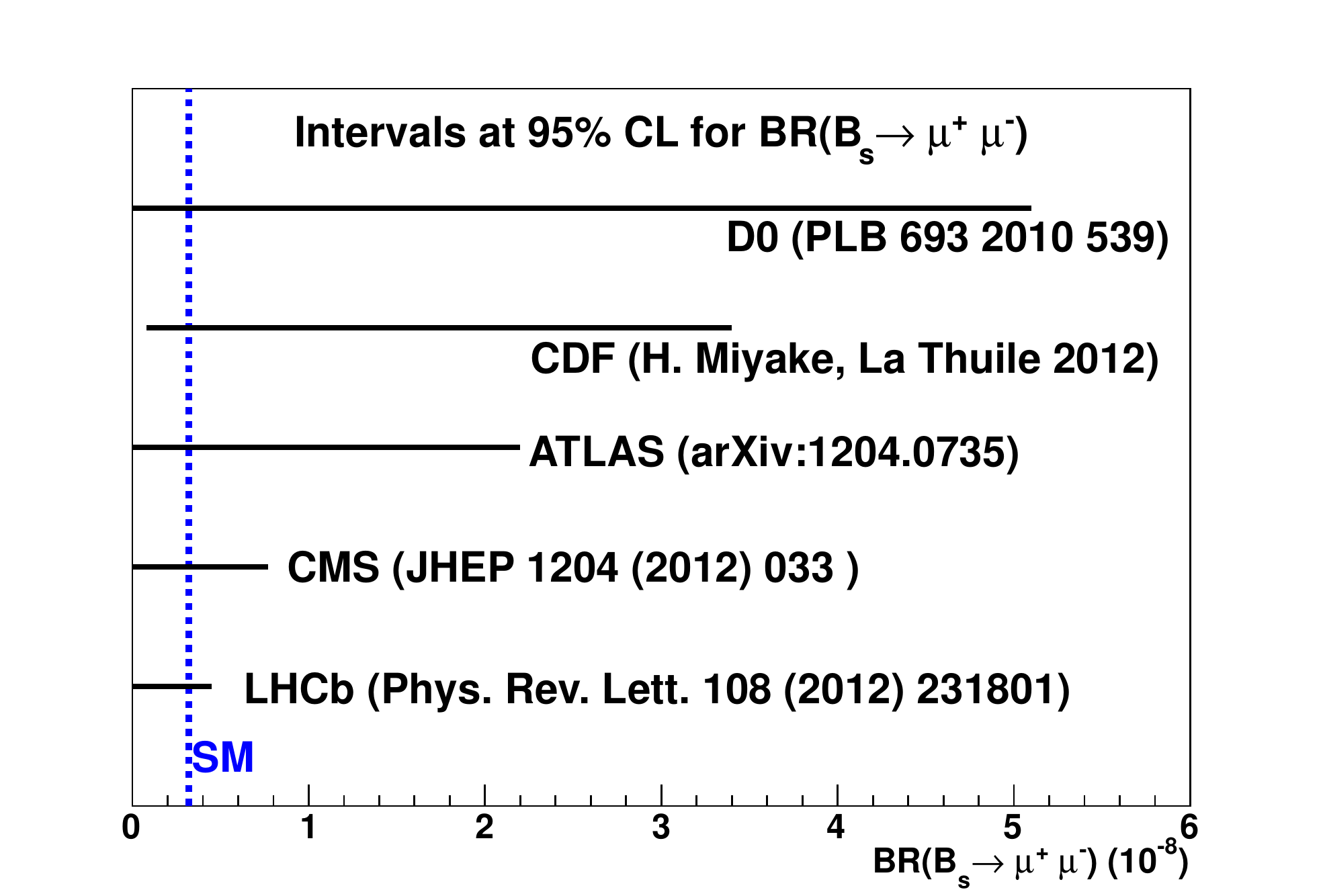}
\caption{Present limits on ${\cal B}(B_s \to \mu^+ \mu^-)$ at $95\%$
  CL set by the experiments D0~\cite{Abazov:2010fs}
  CDF~\cite{Miyake:2012}, ATLAS~\cite{ATLAS-CONF-2012-010}
  CMS~\cite{Chatrchyan:2012rg} and LHCb~\cite{Aaij:2012ac}. The
  SM prediction is indicated by the blue-dashed line. \label{fig:BsmumuInt}}
\end{figure}

In general these branching fractions are given by the following formula:
\begin{equation}
BR(B_{q}\to \mu^+ \mu^-) \propto 
  \left(   \sqrt{1-\frac{4m_{\mu}^2}{m_{B_{q}}}} \mid \frac{m_{B_{q}}^2}{2} ( C_{S}-C_{S}^{\prime} )
    \mid^2 \right) +\left( \mid
    \frac{m_{B_{q}}^2}{2} (C_P - C_P^{\prime}) + m_{\mu} (C_{10} - C_{10}^{\prime})  \mid^2   \right),
\end{equation} 

\noindent where $C_{10}^{(\prime)}$ is an electroweak axial-vector Wilson coefficient and  $C_{S}^{(\prime)}$ and
$C_{P}^{(\prime)}$ are scalar and pseudo-scalar coefficients. In
the SM the (pseudo)-scalar operators are negligibly small, but this 
is not generally the case in NP models. For instance in models with an extended Higgs
sector~\cite{Hurth:2008jc}. 
In particular, in the Minimal Supersymmetric extension
of the SM these branching fractions are proportional to $\tan^6{\beta}$,
where $\tan{\beta}$ is the ratio of the vacuum expectation value of the two Higgs doublets. 
Constraints on these branching fractions  have been set by the
experiments D0~\cite{Abazov:2010fs}, CDF~\cite{Miyake:2012},
ATLAS~\cite{ATLAS-CONF-2012-010}, CMS~\cite{Chatrchyan:2012rg} and
LHCb~\cite{Aaij:2012ac}. The
strongest upper limit was set by LHCb: $BR(B_s\to
\mu^+ \mu^-)<4.5 \cdot 10^{-9} $ and $BR(B_d\to
\mu^+ \mu^-)< 1.0 \cdot 10^{-9}$ at $95\%$CL.
Present constraints for the decay $B_{s} \to \mu^+ \mu^-$ are shown in
Fig.~\ref{fig:BsmumuInt}. \\

\begin{figure}[!h]
\centering
\includegraphics[scale=0.30]{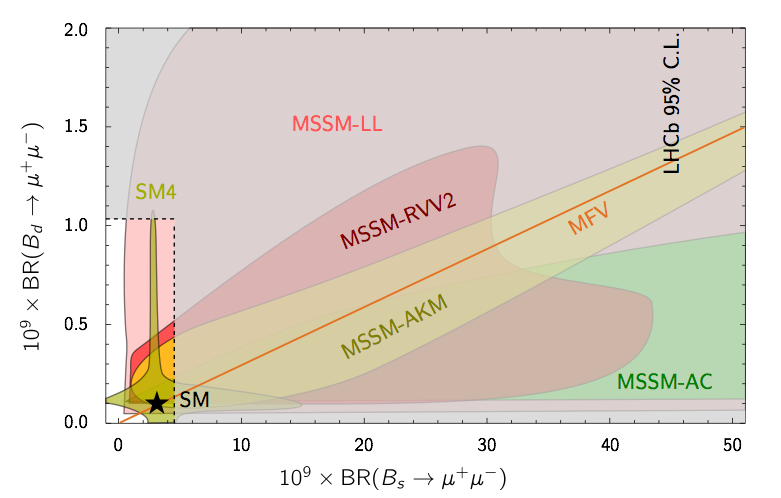}
\caption{Correlation for the branching fractions of the decays $B_s\to
  \mu^+ \mu^-$ and $B_d\to \mu^+ \mu^-$ for several models of NP. Details
  on the models can be found in Ref.~\cite{Straub:2010ih}. The recent upper limits by LHCb are
  shown by the shaded region. Reproduced from Ref.~\cite{Straubb-Moriond}. \label{fig:StraubBsMuMu}}
\end{figure}

At hadron colliders relative measurements of branching fractions
are preferred to absolute ones, in order to minimize systematic uncertainties. 
At the LHCb experiment the three decays $B^+ \rightarrow J/\psi K^+$, 
$B^0\rightarrow K^+ \pi^-$ and $B_s\rightarrow J/\psi \phi$ are used
for the normalization in the measurement of ${\cal BR}(B_s\rightarrow
\mu^+ \mu^-)$\footnote{For the measurement of ${\cal BR}(B_d\rightarrow
\mu^+ \mu^-)$ only the
decays $B^+ \rightarrow J/\psi K^+$, 
$B^0\rightarrow K^+ \pi^-$ are used for normlization.}.  The CDF, ATLAS and CMS experiments use the decay $B^+
\rightarrow J/\psi K^+$ for normalization and CMS also uses the
  decay $B_s\to J/\psi \phi$ as a control channel.  
For all the experiments, the knowledge of the 
fraction of $B_s$ and $B_{d,u}$ mesons produced in the primary $pp$ interaction
$f_s/f_d$ is needed\footnote{Isospin symmetry, i.e.
  $f_u =f_d$, has been assumed.}. This quantity was measured at LHCb by combining
measurements with semi-leptonic and hadronic
decays~\cite{Fleischer:2010ay} to be $f_s/f_d = 0.267^{+0.021}_{-0.020}$~\cite{Aaij:2011jp,Aaij:2011hi}. 
The uncertainty on this parameter is, in the long run, one of the dominant
systematic uncertainties in the measurement of the branching fraction
of the decay $B_s\to
\mu^{+} \mu^{-}$, thereby limiting the discriminating power between SM
and NP contributions in this decay.  Knowledge of $f_s/f_d $ is also
necessary for the measurement of the ratio $\frac{{\cal B} (B_s \to \mu^+ \mu^-)}{{\cal B} 
  (B_d \to \mu^+ \mu^-)}$, which is affected by small theoretical uncertainties
and is a test of Minimal Flavour Violation~\cite{Buras:2010pi}. 
The correlation between the branching fractions of the decay $B_{d}
\to \mu^+ \mu^-$ and $B_s\rightarrow \mu^+ \mu^-$ is shown
in Fig.~\ref{fig:StraubBsMuMu} for several scenarios of physics beyond the SM. 
The recent limits set by LHCb on these decays exclude large fraction of
the parameter space in a whole range of new models.
Large scalar effective couplings are no longer considered a probable
scenario for NP. 
However, it is also possible that NP and SM contributions
interfere destructively, decreasing the $B_{s,d} \rightarrow \mu^+ \mu^-$
branching
fractions. Therefore, precise measurements of  ${\cal B} (B_{s,d} \to
\mu^+ \mu^-)$ remain key observables for the search for physics beyond the SM.  

\section{Angular asymmetries in the decay $B^0 \rightarrow K^{*} \mu^+ \mu^-$}

\begin{figure}[!h]
\centering
\includegraphics[scale=0.4]{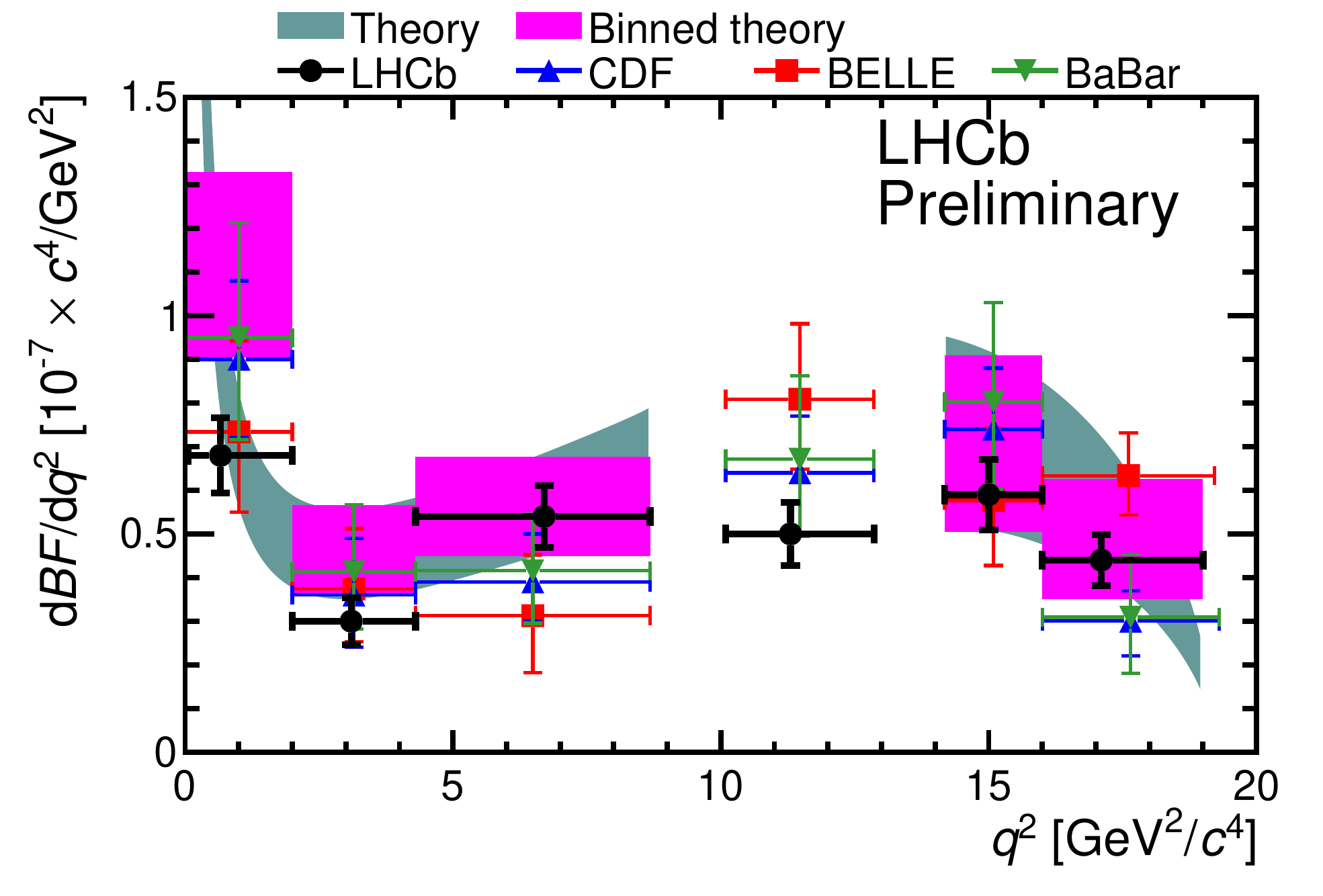}
\caption{$\frac{d{BF}}{dq^2}$ measured by the
  experiments BaBar~\cite{BaBarLakeLouise}, Belle~\cite{Wei:2009zv}, CDF~\cite{Aaltonen:2011ja} and
  LHCb~\cite{LHCb-CONF-2012-008}. The comparison with the SM prediction, taken
  from Ref.~\cite{Bobeth:2010wg,Bobeth:2011nj}, is also shown. Reproduced from Ref.~\cite{LHCb-CONF-2012-008}. \label{fig:Res:dbdq2}}
\end{figure}

\begin{figure}[!h]
\centering
\includegraphics[scale=0.32]{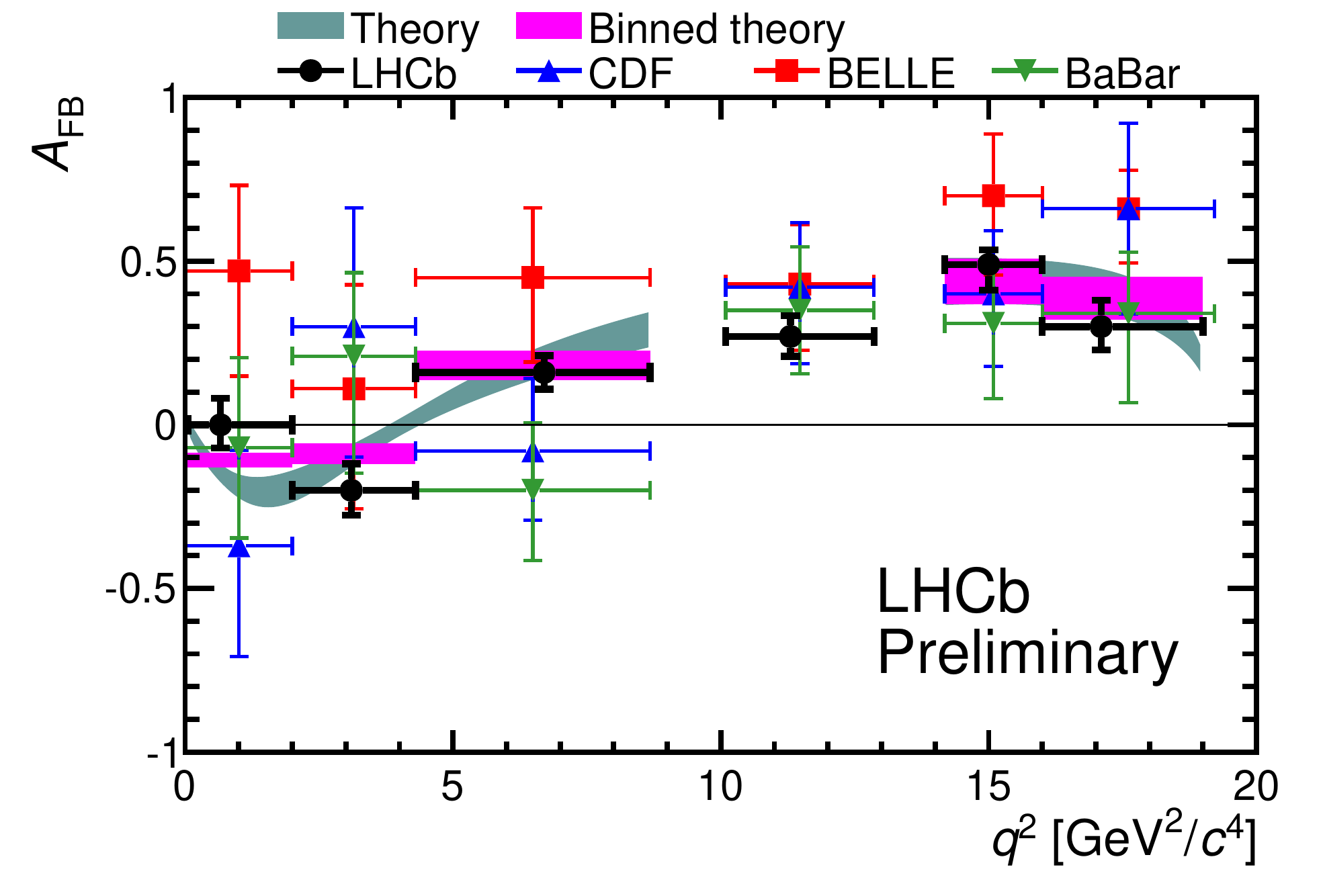}
\includegraphics[scale=0.32]{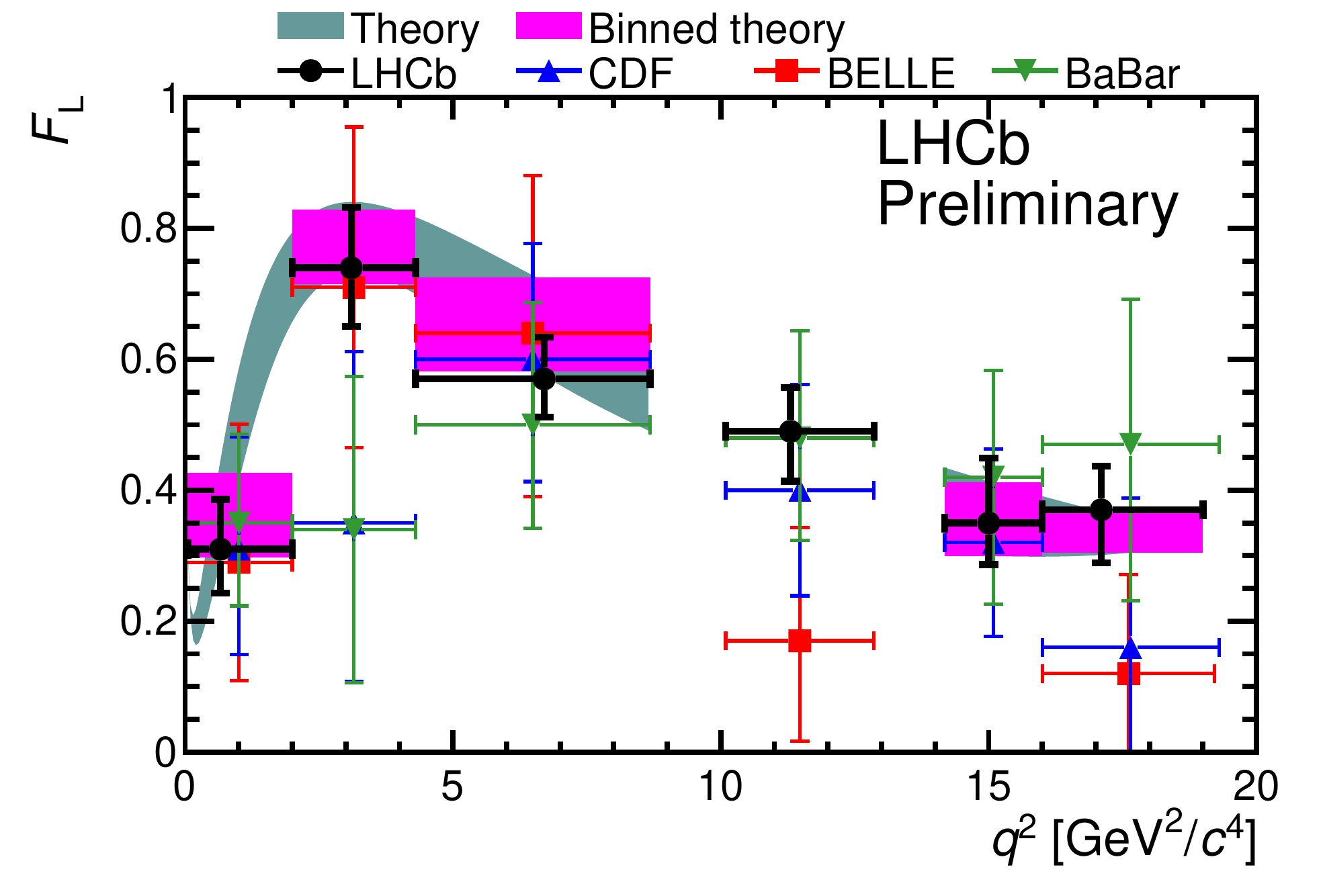}\\
\includegraphics[scale=0.32]{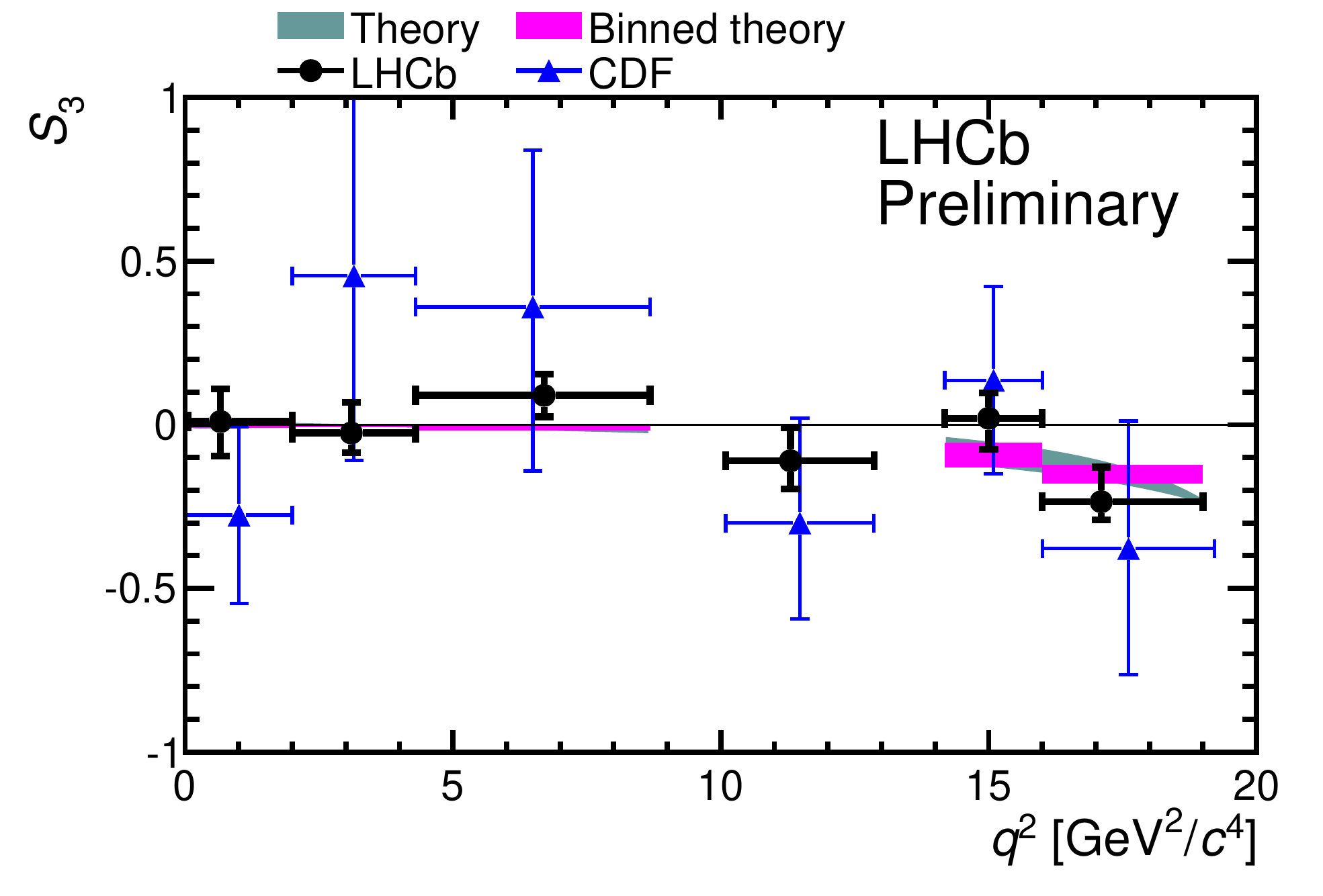}
\includegraphics[scale=0.32]{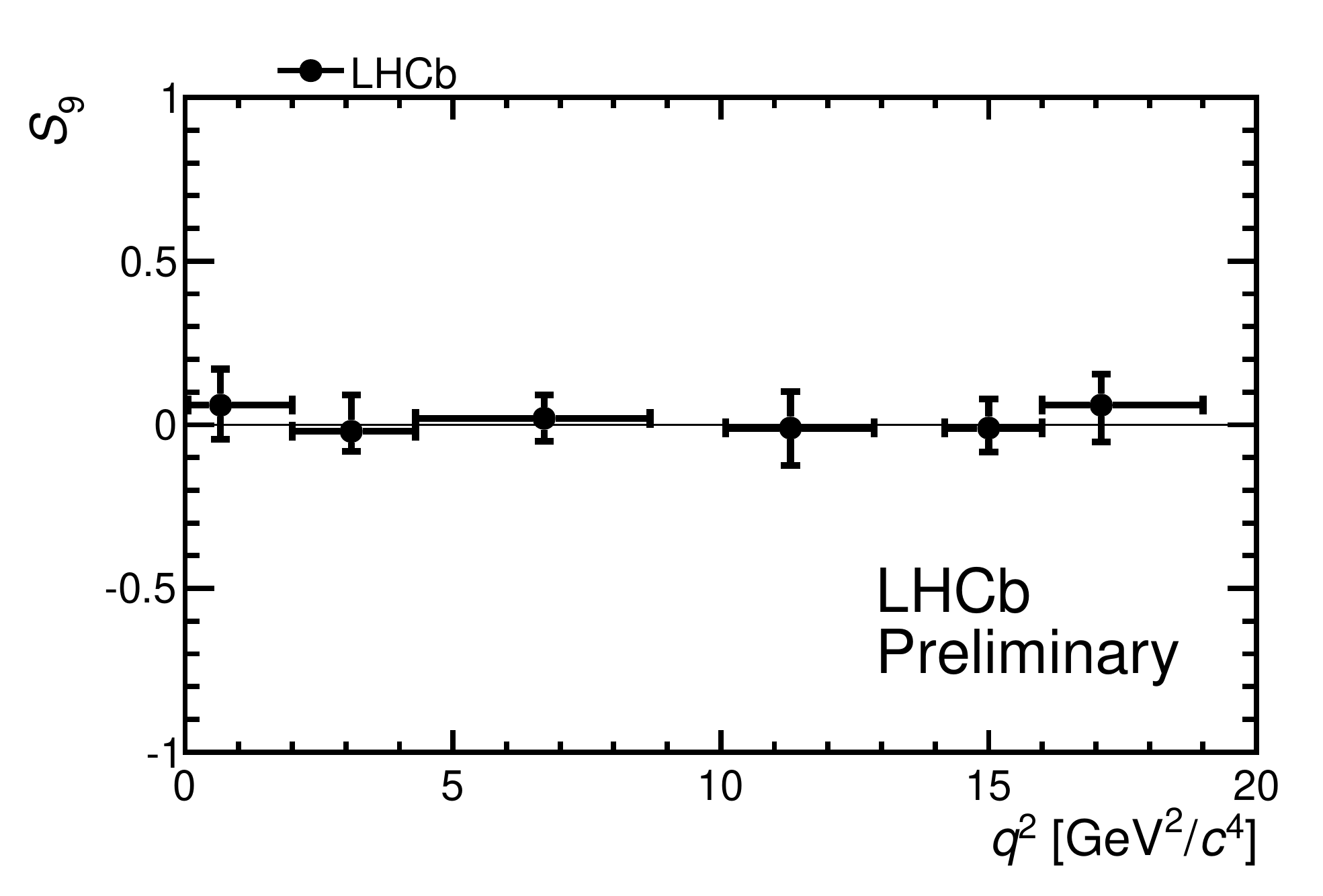}
\caption{$A_{FB}$, $F_L$, $S_3$ and $S_{9}$ measured by the
  experiments BaBar~\cite{BaBarLakeLouise}, Belle~\cite{Wei:2009zv}, CDF~\cite{Aaltonen:2011ja} and
  LHCb~\cite{LHCb-CONF-2012-008}. The comparison with the SM prediction, taken
  from Ref.~\cite{Bobeth:2010wg,Bobeth:2011nj}, is also shown. Reproduced from ~\cite{LHCb-CONF-2012-008}. \label{fig:Res:Kstmm}}
\end{figure}

In the decay $B_d\rightarrow K^* \mu^+ \mu^-$ several angular
observables, which are sensitive to physics beyond the SM, can be
constructed. For some of these observables theoretical uncertainties are under
control or cancel out (see~\cite{Ali:1991is,Altmannshofer:2008dz,Matias:2012xw} and references therein). 
These observables include the forward-backward asymmetry of the dimuon system, $A_{FB}$, the fraction 
of $K^{*}$ longitudinal polarization, $F_L$, the transverse
asymmetry, $S_{3}$ ~\cite{Altmannshofer:2008dz} (often referred to as
$\frac{1}{2}(1-F_L)A_{T}^2$ in the literature~\cite{Kruger:2005ep}),
and the CP averaged quantity $S_{9}$~\cite{Altmannshofer:2008dz}, proportional
to the imaginary component of the product of the transverse and longitudinal amplitude of the $K^{*0}$. \\
These observables can be extracted by performing an angular analysis as a function of the
dimuon invariant mass squared, $q^2$, with
respect to the following angles: the angle $\theta_l$ between
the $\mu^+$ ($\mu^-$) and the $B^0$ ($\overline{B}^0$) in the dimuon
rest frame; the angle $\theta_K$ between the kaon and $B^0$
in the $K^*$ rest frame; and the  angle $\phi$ between the planes of
the dimuon system and the plane of the $K^*$. A formal definition of
these angles can be found in Ref.~\cite{Egede:1048970}. It should be
noticed that the definition of the angles varies in the literature. 
In particular, the sign of the $\phi$ angle in LHCb is opposite to that of
CDF for the $\overline{B}^{0}$ decay, while it is the same for
  the $B^0$ decay. Consequently, in
place of $A_{Im}$ ~\cite{Egede:2008uy} the angular distribution
measured at LHCb is sensitive to the CP-average $S_{9}$, whereas
 CDF is sensitive to the CP asymmetry $A_{9}$~\cite{Altmannshofer:2008dz,Bobeth:2008ij}. Recent measurements of these observables are shown in
Fig.~\ref{fig:Res:Kstmm}. These measurements are dominated by the LHCb
results~\cite{LHCb-CONF-2012-008}. 
Measurements of the differential branching fraction as function of $q^2$
are shown in Fig.~\ref{fig:Res:dbdq2}. All these measurements are
consistent with each other and with SM predictions.

The point in $q^2$ where $A_{FB}$ changes sign
is a sensitive probe for NP and it is theoretically clean, since form
factor uncertainties cancel out at first order. 
The LHCb collaboration reported the world's first measurement of this
observable, shown in Fig.~\ref{fig:Res:Kstmm}. The measurement of the
zero-crossing point is $q^2_0 =4.9^{+1.1}_{-1.3}$~GeV$^2$/c$^4$, which
is in agreement with the SM prediction. This measurement strongly
disfavours scenarios with flipped $C_7$ sign with respect to the SM,
which early measurements of $A_{FB}$ seemed to be hinting at. \linebreak

\section{Isospin asymmetry in the decays $B\to K^{(*)}l^+l^-$}
\begin{figure}[!h]
\centering
\includegraphics[scale=0.6]{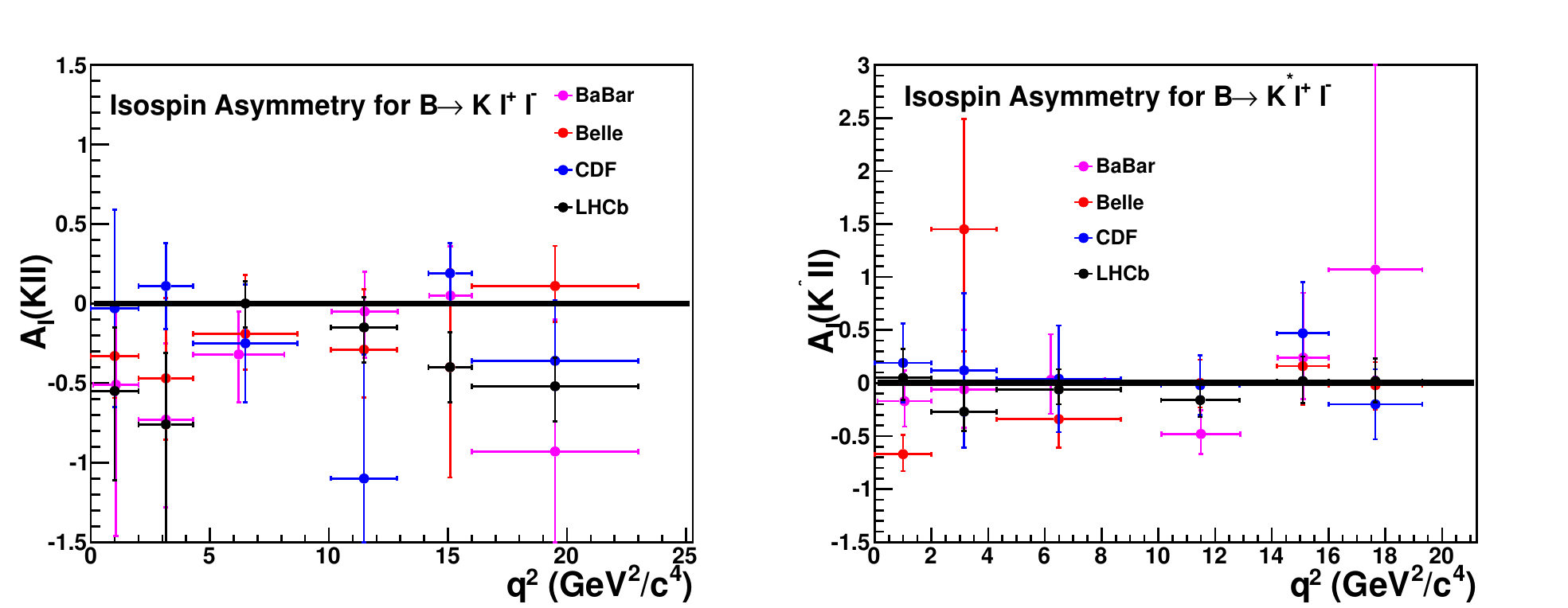}
\caption{Isospin asymmetry for the decays $B\to K^{( * )} l^{+} l^{-}$,
  measured by the experiments BaBar~\cite{BaBarLakeLouise},
  Belle~\cite{Wei:2009zv} (with electrons and muons) , 
  CDF~\cite{Aaltonen:2011ja} and LHCb~\cite{Aaij:2012cq} (with muons).\label{fig:Isospin}}
\end{figure}

The isospin asymmetry
of the decays $B\to K l^+l^-$ and $B\to K^* \mu^+ \mu^-$ are defined as:
\begin{equation}
A_I = \frac{{\cal B}(B^0\to K^{(*)0} l^+
  l^-)-\frac{\tau_0}{\tau_+}{\cal B}(B^{\pm}\to K^{(*)\pm} l^+
  l^-)}{{\cal B}(B^0\to K^{(*)0} l^+ l^-)+\frac{\tau_0}{\tau_+}{\cal
    B}(B^{\pm}\to K^{(*)\pm} l^+ l^-)},
\end{equation} 
where $\tau_{0,+}$ is the $B^{0,+}$ lifetime. In the SM, this quantity is expected to
be at the percent level. 
Measurements of this quantity have been
made by the BaBar~\cite{BaBarLakeLouise} and Belle~\cite{Wei:2009zv} experiments, using electrons and
muons and by CDF~\cite{Aaltonen:2011ja} and LHCb~\cite{Aaij:2012cq} , using muons. These results are shown in
Fig.~\ref{fig:Isospin}.  All the measurements are consistent with each other and
they are consistent with SM predictions for the $B\to K^{*}l^+l^-$
decays. For the $B\to K l^+l^-$ decays the measurements are in
agreement with each other but they show a tension with respect to
naive expectations. In particular, LHCb data show a deviation
from zero at the level of about four standard deviations~\cite{Aaij:2012cq}. At present there is no theoretical explanation for this large isospin
asymmetry.  

\section{Conclusions}

Measurements in flavour physics have a great track record in paving the way to significant discoveries in particle physics. 
Most NP scenarios predict deviations from SM expectations in rare
B-meson decays. Sensitive probes for NP are the leptonic decays
$B_{s,d} \to \mu^+ \mu^-$ and the rare semi-leptonic decays $B\to
K^{*}l^+ l^-$. 
The most recent measurements of the decays $B_{s,d} \to \mu^+ \mu^-$
and the decay $B_d\to K^* \mu^+ \mu^-$ are in good agreement with SM
predictions and set strong constraints on a number of NP models. 
The isospin asymmetry in the decays $B \to K^{(*)} l^+ l^-$ has been
measured by several experiments. These measurements agree with each
other and with SM predictions for the decays with a $K^{*}$, while
there is a significant tension with respect to expectations for the decays with a kaon. 

While no convincing sign of NP has been found yet in rare decays,
these measurements are at the moment statistically limited and 
the room for physics beyond the SM is still large. Future and more precise measurements will
be important to test SM predictions and understand the flavour structure
of NP. 

\begin{acknowledgments}
The author would like to thank the conference organisers for
the warm hospitality.   
\end{acknowledgments}

\end{document}